\documentclass[%
 aip,
 amsmath,amssymb,
 reprint,%
]{revtex4-1}

\usepackage{graphicx}% Include figure files
\usepackage{dcolumn}% Align table columns on decimal point
\usepackage{bm}% bold math
\usepackage[utf8]{inputenc}
\usepackage[T1]{fontenc}
\usepackage{mathptmx}
\usepackage{etoolbox}
\usepackage{xcolor}
\makeatletter
\def\@email#1#2{%
 \endgroup
 \patchcmd{\titleblock@produce}
  {\frontmatter@RRAPformat}
  {\frontmatter@RRAPformat{\produce@RRAP{*#1\href{mailto:#2}{#2}}}\frontmatter@RRAPformat}
  {}{}
}%
\makeatother

\begin{document}

\preprint{AIP/123-QED}

\title[]{Design of a test rig for the investigation of falling film flows with counter-current gas flows}
% Force line breaks with \\
\author{M. Wirth}
 \email{Markus.Wirth@itlr.uni-stuttgart.de}
 \affiliation{Institute of Aerospace Thermodynamics, University of Stuttgart, Pfaffenwaldring 31, 70569 Stuttgart, Germany}
 \author{J. Hagedorn}

\affiliation{%
Institute of Thermodynamics, Leibniz University Hannover, An der Universität 1,30823 Garbsen, Germany
}%
\author{B. Weigand}%
  \affiliation{Institute of Aerospace Thermodynamics, University of Stuttgart, Pfaffenwaldring 31, 70569 Stuttgart, Germany}
 \author{S. Kabelac}

\affiliation{%
Institute of Thermodynamics, Leibniz University Hannover, An der Universität 1,30823 Garbsen, Germany
}%

\date{\today}% It is always \today, today,

\begin{abstract}
Geothermal phase change probes operate on the principle of falling film evaporation, enabling the efficient use of geothermal heat for space heating applications. Despite successful applications in research, their commercial use is limited. One of the primary reasons for this is the absence of validated models capable of accurately representing the falling film flow coupled with counter-current gas flow within these probes. For this reason, a test rig has been developed to facilitate the validation of future models. This test rig allows for the replication of flow phenomena appearing within geothermal probes under controlled conditions. The test rig was successfully tested within the first measurement campaign presented in this paper. Within the framework of this measurement campaign water was used as liquid due to its similar Kapitza number compared to \(CO_{2}\), typically used within geothermal probes, and humid air as gas. For the gas phase, velocity profiles were measured and for the liquid phase high temporal resolution film thickness measurements were conducted. Both measurement systems show qualitative comparable results to  literature. Additionally, the sampled film thickness data were averaged to enable a time-independent interpretation. 
The investigation of the average film thickness in flow direction, without gas flow, revealed an increase in film thickness along flow direction for \(Re_{film}=500\). In contrast, for \(Re_{film}\geq 980\), an opposing trend was observed. Furthermore, the influence of gas flow on the average film thickness was investigated. The measurement results and high-speed camera images indicate that the test rig is capable of reaching flooding conditions.
\end{abstract}

\maketitle

\section{\label{sec:level1}Introduction}

\begin{figure}[!ht]             % Einbetten in figure wie gehabt
\centering                  % zentrierte Ausrichtung, optional
\def\svgwidth{270pt}    
% die Bildbreite muss auf diese Weise festgelegt werden!
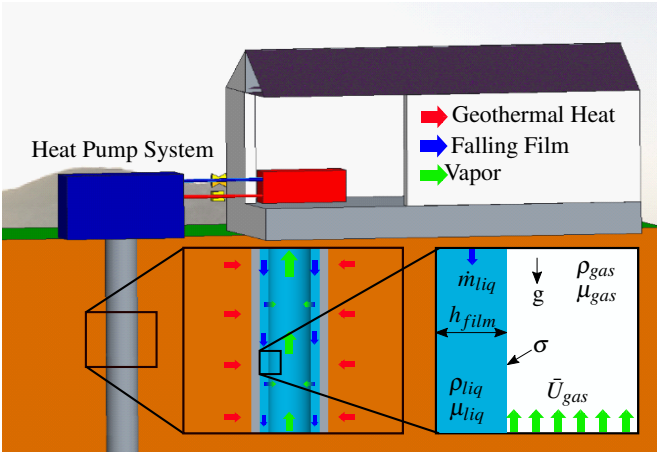  
% 'versuchsaufbau' durch Dateinamen ersetzen.
\caption{Working principle of a TCPT for space heating application}    % Bildunterschrift, optional
\label{fig:Thermosyphon}          % Label f�r Verweise, optional
\end{figure}

Falling film flows play an important role in numerous technological applications like desalination \cite{MACIVER2005221}, distillation \cite{BATTISTI2020107873}, absorption \cite{NARVAEZROMO20171079}, chemical processing or heat recovery \cite{NA2020101810}. Applications where falling films are typically applied, require high heat and mass transfer rates while temperature differences are small due to the thermal sensitivity of the process \cite{ZHAO2022117869,BATTISTI2020107873}, or the lack of high temperature differences availability \cite{NA2020101810}. The latter condition is applicable to near-surface geothermal heat pumps, where extracted geothermal heat is employed as a heat source for residential space heating. In such applications, the temperature available from the soil typically ranges from 5°C to 20°C  \cite{Hagedorn_2024}, remaining relatively constant throughout the year. Due to the low temporal temperature fluctuations, this heat source enables heat pumps to operate at higher energy efficiencies when compared to conventional air heat pumps \cite{RUI2018450,BLOOMQUIST2003513}. To utilize the beneficial properties of falling film flows for geothermal heat extraction, so called two-phase closed thermosyphons (TPCT) are used. Such a TPCT, is illustrated in Fig.\ref{fig:Thermosyphon}, for the purpose of space heating. The TPCT consists of a closed pipe embedded within a borehole reaching lengths up to 400 m \cite{Hagedorn_2024}. The pipe is filled with a working fluid under two-phase conditions. The TCPTs top is connected to a heat pump system, responsible for the heat extraction. The heat transfer causes the gaseous phase of the working fluid to condensate. The condensed fluid accumulates to a liquid film flowing downstream within the closed pipe. The geothermal heat is transferred from the soil into the falling film. This results in a continuous evaporation of the falling film in downstream direction.  The vapor arises to the top of the probe where the latent heat is transferred to a heat pump causing condensation and thus, film reformation. In comparison to commonly used single-phase U-shaped probes utilizing brine or a water/glycol mixture as a working fluid, TCPTs extract 15-25\%  more geothermal energy at the same electrical power input\cite{Storch2018} . This enhancement relies on the working principle, obviating the necessity for fluid pumps and reducing conductive thermal losses. TCPTs were already investigated in large scale experiments with the working fluids propane \cite{StorchDiss} and CO$_{2}$  with plane \cite{Hagedorn_2024} as well as corrugated pipes \cite{Ebeling2017,Kruse2010}. Despite the advantage of these systems, their commercial use is limited due to a lack of validated models available, hindering design and optimization efforts. This apparent lack of models can be attributed to the complex flow phenomena within such probes. These phenomena encompass not only the film flow but also the counter-current vapor flow and the liquid-vapor interaction. Due to size, inaccessibility and dynamic behavior of TCPTs, investigations of the internal phenomena are limited, thus unsuitable for determining desired models. Consequently, the investigation of the hydrodynamic phenomena under controlled conditions is essential for future model validation and improvement.
The first modelling approach of a falling liquid film  was developed by Nusselt \cite{Nusselt}. The liquid was assumed as a thin laminar film with constant material properties, a shear-stress free, smooth interface. Based on this assumptions the film height can be derived as 
\begin{equation}
 h_{film}=(3 \frac{\mu_{liq}^2}{\rho_{liq}^{2}g})^{1/3}Re_{liq}^{1/3}%
\label{eq:three}
\end{equation}
with the gravitational acceleration \(g\) , the film Reynolds number \(Re_{film}\), the dynamic viscosity \(\mu_{liq}\) and density \(\rho_{liq}\) of the liquid. In this work, the film Reynolds number of the falling film is defined as
\begin{equation}
 Re_{liq}= \frac{\dot{m}_{liq}}{L \mu_{liq}}%
\label{eq:one}
\end{equation}
where   \(\dot{m}_{liq}\) is the liquid mass flow rate and \(\ L\) is defined as the wetted unit length. The Reynolds number is often used to distinguish between different hydrodynamic flow regimes. In case of falling film flows using only \(Re_{film}\) is insufficient for predicting these regimes. According to Al-Sibai \cite{Al-Sibai_diss} the Kapitza number
\begin{equation}
\ Ka = \frac{\rho_{liq} \sigma^{3}}{\mu_{liq}^{4} g }%
\label{eq:two}
\end{equation}
with the surface tension \(\sigma\) must be considered for a correct flow regime prediction. In case of the gas flow, the Reynolds number is defined as  
\begin{equation}
 Re_{gas}= \frac{D_{h} \overline{U}_{gas} \rho_{gas}}{\mu_{gas}}%
\label{eq:three}
\end{equation}
where \(D_{h}\) is the hydraulic diameter and \(\overline{U}_{gas}\) is the superficial gas velocity. Numerous studies were conducted on counter-current film-gas flows within circular pipes. However, the design of these pipes presents challenges related to low accessibility due to their full circumferential wetting. Consequently, detailed investigations of the flow phenomena are only possible to a limited extent. As determined by Kapitza \cite{Kapitza}, the pipe curvature is negligible with a large ratio of pipe diameter to film thickness. This condition applies to TCPTs, as estimated from the calculated data provided by Ebeling et al. \cite{Ebeling2017}. This allows observations of similar flow phenomena on plane walls, leading to a considerable improvement in accessibility. In contrast to pipe flows, investigations of falling film flows with counter-current gas flow on flat planes have been conducted less frequently. Drossos et al. \cite{DROSOS200651} investigated the influence of the gas flow on the instantaneous, local film thickness and wall shear stress in a vertical rectangular channel. Water and water-butanol solutions were used as liquids for \(Re_{liq}\) < 350 and \(\overline{U}_{gas}\)  < m/s . Huang et al.  \cite{HUANG2014125} conducted water-air counter-current flow investigations within a large scale facility featuring an inclineable plate with a length of 5m. Temporal resolved, local film thickness measurements were made for inclination angles of 30° and 60° for \(100<Re_{liq}<1500\) and \(\overline{U}_{gas}\) < 15 m/s. Al Sayegh et al. \cite{SAYEGH2022104170} investigated the spatial and temporal thickness distribution on inclined plates and corrugated plates with an optical technique based on fluorescence intensity. Water-butanol-fluorescein was used as liquid in a range from \(13<Re_{liq}<290\), \(\overline{U}_{gas}\) < 5m/s, and inclination angels varying from 10° to 60°. Unfortunately, none of the above mentioned facilities can be used to replicate the hydrodynamic conditions within TCPTs due to too low Reynolds numbers\cite{DROSOS200651}, insufficient Kapitza numbers \cite{HUANG2014125},or both \cite{SAYEGH2022104170}.
To adress this gap , a test rig called Falling Film Evaporation Test-Rig (FETRIG) has been developed at the Institute of Aerospace Thermodynamics (ITLR) and is presented in this study. FETRIG has been designed for the investigation of falling film and counter-current gas flows under highly controlled conditions, allowing the reproduction of various operating points within TCPTs. The investigations are conducted within a measurement section characterized by its modular design and optical accessibility.  This design enables the utilization of measurement techniques that are otherwise restricted by accessibility limitations in TCPTs.  This paper first shows a description of the experimental facility and the measurement methods, followed by qualitative and quantitative results of the first measurement campaign. Within the framework of this measurement campaign, water was used as liquid due to its comparable Kapitza number to \(CO_{2}\) and humid air. Velocity profiles of the gas phase were measured, while high temporal resolution film thickness measurements were conducted for the liquid phase. Additionally, high-speed camera imaging of the film was performed to obtain qualitative insights.

\subsection{\label{sec:level2}Test rig for falling film investigations}

\begin{figure*}[htbp] % Use figure* to span columns
    \centering
    \includegraphics[width=\textwidth]{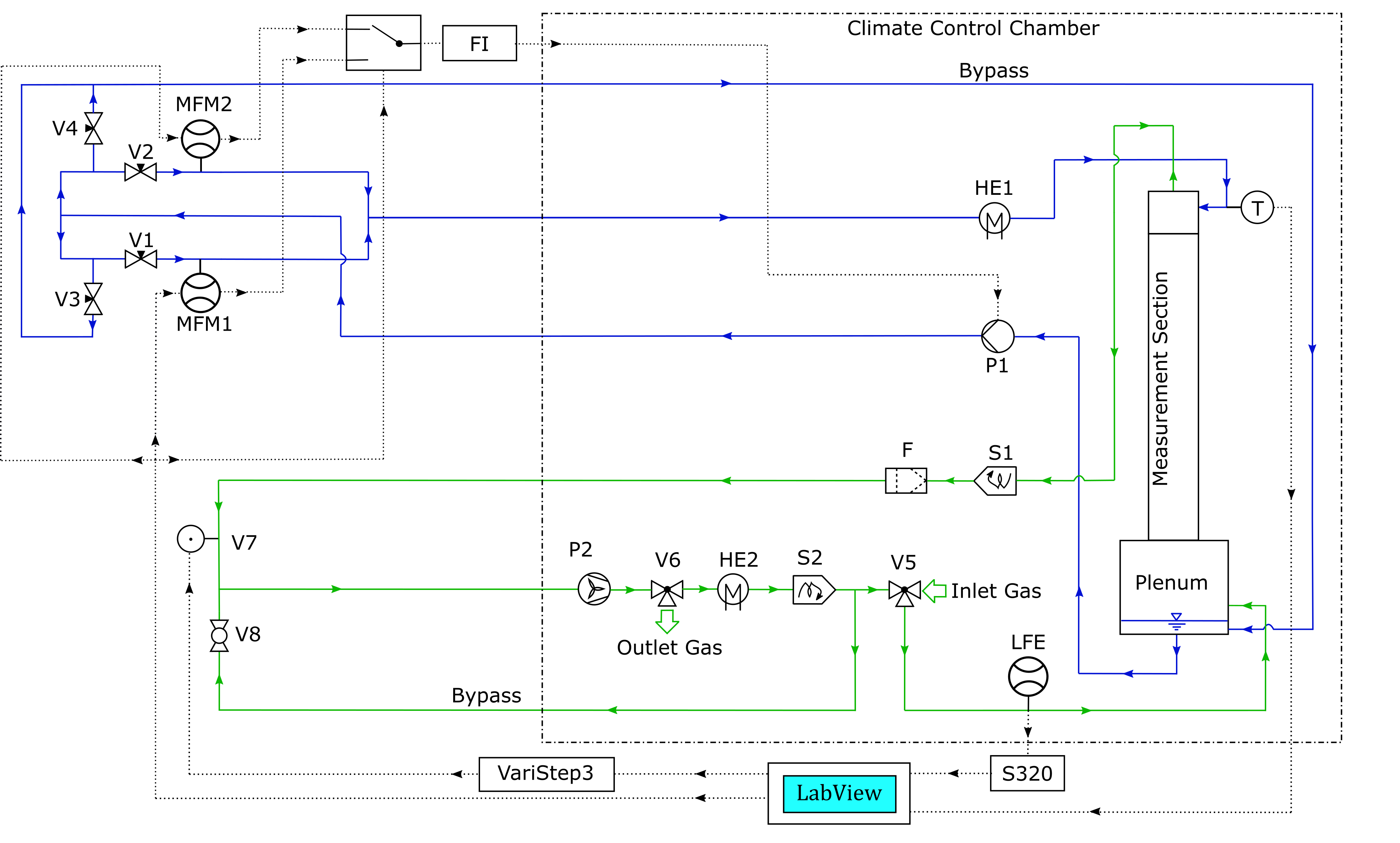} % Adjust the width to your needs
    \caption{Schematics of the Falling Film Evaporation Test Rig (FETRIG) with the liquid circuit (blue), the gas circuit(green) and data lines (dotted lines).}
    \label{fig:FETRIG}
\end{figure*}

A schematic drawing of FETRIG is shown in Fig.\ref{fig:FETRIG}.  FETRIG consists of the measurement section, a liquid circuit (blue) and a gas circuit (yellow). The measurement section is located within a climate control chamber, allowing temperature and humidity variations ranging form -10°C to 60°C and 5 to 95 \% relative humidity. This ensures constant conditions during measurements, repeatability between measurements and variations of the fluids material properties if desired. Components that are sensitive to such conditions or must be accessed during measurements are located outside the climate chamber. 
All of FETRIGs electronic measurement and control devices  are embedded within a single LabVIEW environment. This environment ensures monitoring and recording of the measurement data including temperatures, pressures, humidity  and flow rates of the gas and liquid during measurements. Additionally, the framework enables the independent regulation of the gas and liquid Reynolds numbers. The configurations of the circuits for this purpose, along with their operational principles, are described in greater detail in the following sections.

\subsubsection{\label{sec:level3} Liquid circuit and control}
For the liquid circuit, the liquid stored within the measurements section plenum is transported by the peripheral Pump P, specifically a Fink Chem+Tec type F-01-900-5. It is directed to one of two mass flow meters: MFM1, a Bronkhorst Coriflow M55-AGD-HH-0-S, and MFM2, a Bronkhorst Coriflow M14-AGD-HH-0-S. Both mass flow meters are based on the Coriolis principle, which allows high resolved independent measurements of fluid mass flow rates and fluid densities. Additionaly, MFM1 and MFM2 are equipped with internal PID controllers for optional mass flow control by means of separately mounted control valves or pumps. The selection between MFM1 and MFM2 is done via the valves V1 and V2 and depends on the desired massflow to be investigated. With MFM1 covering a measurement range of $2.78\times 10^{-5}$-$2.78\times10^{-3}$ kg/s and MFM2 of $1\times10^{-3}$-0.1 kg/s. After passing the corresponding  MFM the liquid gets transferred through a liquid heat exchanger for temperature adjustment before re-entering the measurement section at the liquid inlet. 
The desired liquid Reynolds number is set within FETRIGs LabVIEW environment. By using the liquid temperature at the liquid inlet of the measurement section T, the dynamic viscosity at the inlet of the measurement section is calculated within the LabVIEW environment. According to Eqs.~(\ref{eq:two}) the desired mass flow rate  is determined and transmitted to  MFM1 and MFM2.  The corresponding PID signal is calculated with the target value and the measured mass flow. The selection of the PID signal to be used is determined in LabVIEW based on the size of the target value and the corresponding resolution of the MFMs. The selected PID signal is then transmitted to the pump's frequency inverter (FI). The FI converts this signal into the desired rotational speed of the pump, resulting in the corresponding mass flow. If the PID control does not achieve the desired mass flow rate, the control range can be fine-adjusted using the needle valves V3 or V4 of the bypass line. 

\subsubsection{\label{sec:level3} Gas circuit and control}
In contrast to the liquid circuit the gas circuit can be operated in two different modes, the open configuration mode and the closed configuration mode. In the open configuration mode conditioned air within the climate chamber is sucked into the circuit and exhausted after passing the measurement section, control devices and the blower P2 into the climate chamber again. This mode is used for hydrodynamic investigations of water or aqueous solutions. In such cases the humidity inside the climate chamber is set to maximum humidity to inhibit undesired evaporation while maintaining well defined inlet temperatures for isothermal investigations. In the closed configuration mode, in contrast, the gas is recirculated and conditioned within a closed loop. This configuration is employed during evaporation experiments, where contamination of the working fluids with ambient air is undesired, and the release of the working fluid vapor must be avoided. The adjustment of the configuration mode is done via the two three way valves V5 and V6.  The subsequent description of the circuit is based on the open configuration mode. The conditioned air enters the circuit at V5 and passes through the Laminar Flow Element (LFE), specifically a Terta Tec, 50MC, which is used for volume flow measurement. For this purpose, the pressure loss is measured across the LFE, as well as the absolute pressure and gas temperature at the LFEs inlet and the ambient relative and absolute humidity. The measured values are used as input for the calculation of the volume flow rate and viscosity that takes place within the corresponding LFE controller, Tetra Tec S320. This controller is calibrated for volume flows of up to 0.05 m³/s. Both, the caluculated values as well as the input values are transmitted to the LabVIEW environment. Within the environment the measured gas Reynolds number  is calculated according to Eq.~(\ref{eq:three}).  After the LFE, the gas passes the measurement section. The gas exiting the measurement section then passes a cyclone seperator S1 followed by a gas filter F.  Consequently, liquid entrained from the film within the measurement section is separated from the gas flow to prevent unwanted liquid dispersion within the gas circuit, which could potentially cause harm to components. The purified gas subsequently flows through the electrical throttle valve V7 ( Motortech, ITB 50)  and the blower P2 used for volume flow control. For flow control, the set and measured gas Reynolds number is utilized in the calculation of the PID control signal within the LabVIEW environment.  The control signal is transmitted to a Motortech, VariStep3 throttle valve driver, where it is transduced into a corresponding throttle flap angle. This alternation of the flap angle modifies the pressure drop characteristics of the circuit, which in turn shifts the operational point of P2, leading to a variation in gas flow rate. Upon exiting the blower the gas either exits the circuit at V6 or is redirected through the gas heat exchanger HE2 and cyclone separator S2, returning to the inlet of P2. This bypass line is controlled manually via the membrane valve V8 and is used for adjustments of the circuits control range. In case of the closed circuit configuration the heat exchanger HE2 is used for adjusting the operational gas temperature. In case of film evaporation,  the heat induced to the system is extracted by HE2. The extraction, therefore, is causing re-condensing of the working fluid. Subsequently, the liquid is segregated by separator S2 to prevent potential damage to the LFE.

\subsection{\label{sec:level2}Measurement section}
\begin{figure}[h!]
  \includegraphics[width=1\columnwidth]{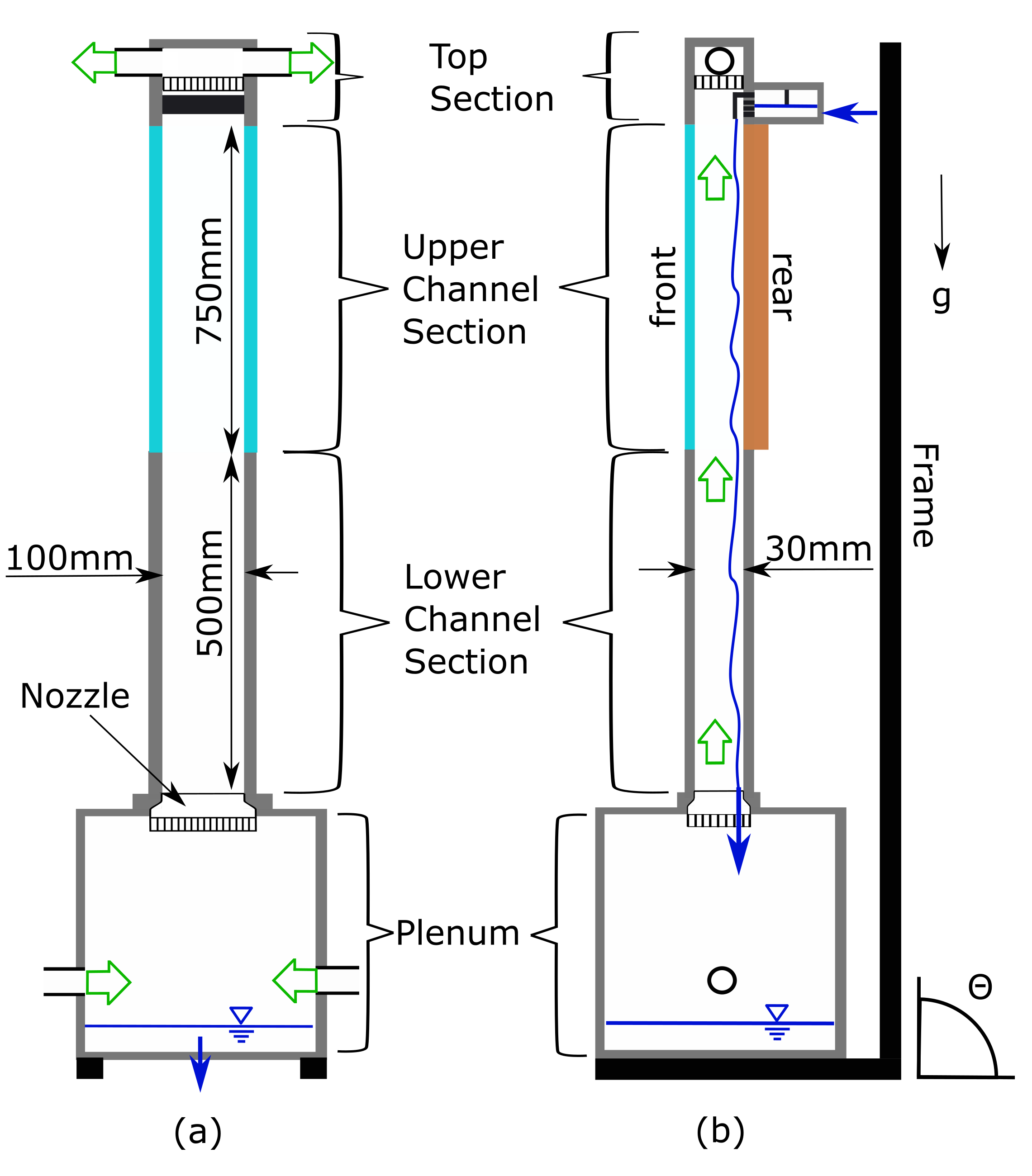}
  \caption{Schematics of the measurement section with (a) frontal section view and (b) side section view}
  \label{fig:Measurement_Section}
\end{figure}

Figure \ref{fig:Measurement_Section} shows a schematic illustration of the measurement section. 
The measurement section consists of the top section, upper channel section, lower channel section and a plenum. The upper section has a length of 750 mm, while the lower channel section extends 500 mm in length. Both are designed with a rectangular cross-section, measuring 100 mm in width and 30 mm in height. The liquid film flows on the wall on the backside of the measurement section. This geometric configuration ensures optimal accessibility of the film and straightforward integration of measurement devices. Additionally, an increase in modularity is achieved due to the rectangular design. 
The measurement section is integrated into an adjustable hanging, frame allowing investigations at inclination angles ranging from 60° to 90°. 

\begin{figure}[h!]
  \includegraphics[width=1\columnwidth]{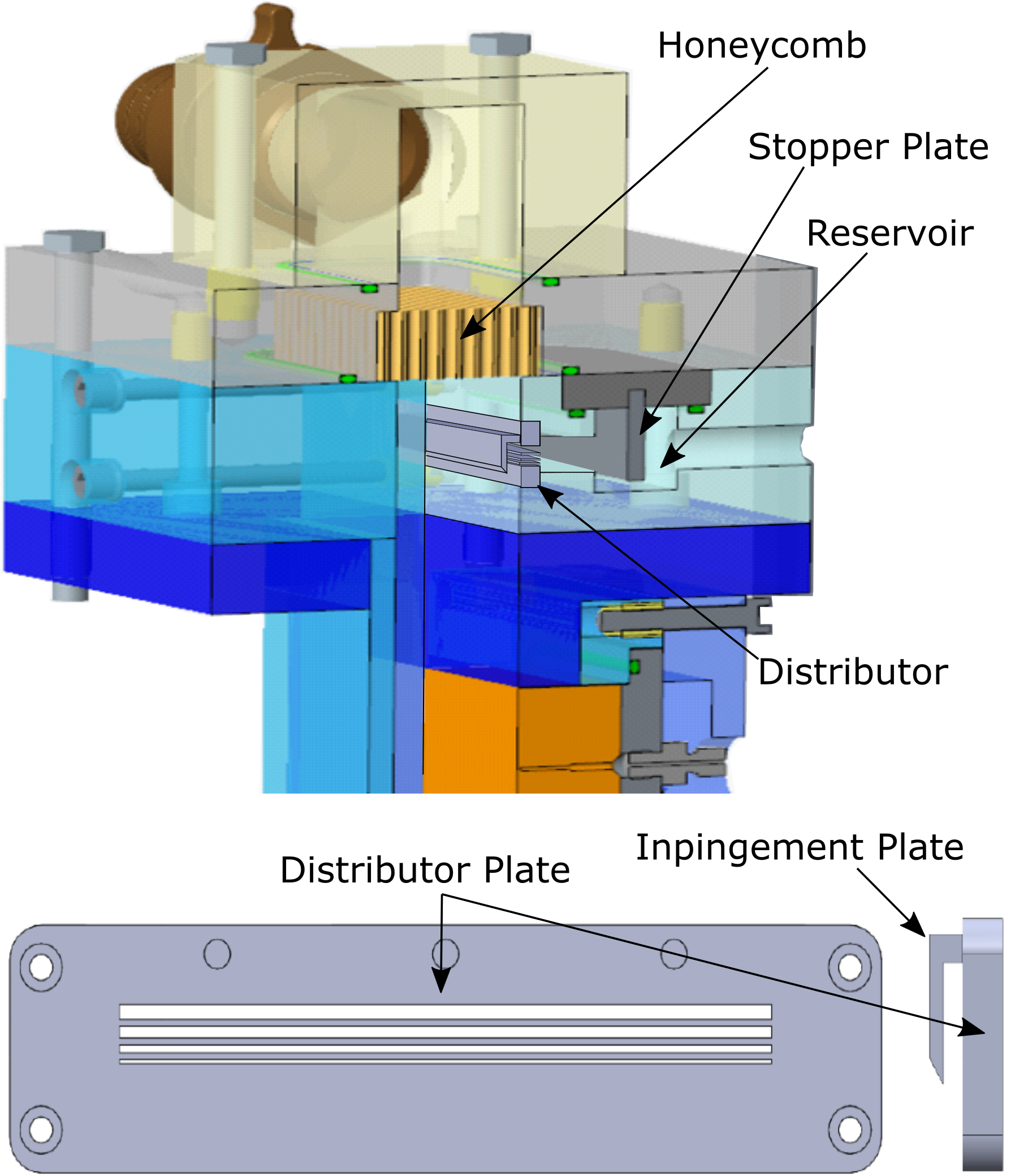}
  \caption{Cross section view of the top section and film distributor}
  \label{fig:Top_Section}
\end{figure}

\begin{figure}[h!]
  \includegraphics[width=1\columnwidth]{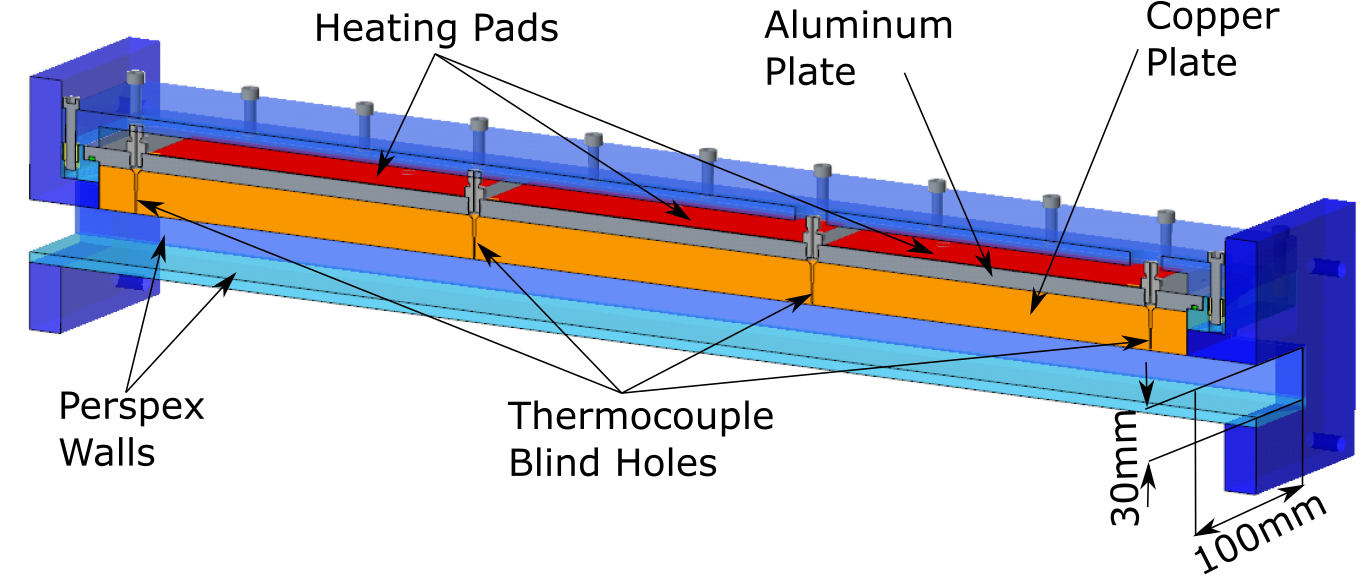}
  \caption{Sectional view of the upper channel section}
  \label{fig:Channel}
\end{figure}

Within the top section, the gas exits the measurement section and the liquid film is generated. A cross-sectional view of the top section is displayed in Fig \ref{fig:Top_Section}. The two gas outlets are positioned on the side walls of the channel, due to the limited height of the climate chamber. To reduce the influence of streamline curvature of the gas main flow, caused by this design, a honeycomb structure is placed in front of the outlets. For film generation, the liquid is introduced through inlets located on the backside, passes through an internal reservoir, that decelerates the flow,  and subsequently enters the channel via the film distributor. Additionally, an interchangeable stopping plate is placed inside the reservoir.  As a consequence, a direct impact of the liquid onto the distributor plate is avoided, leading to an uniform liquid supply for film generation. The interchangeable film distributor consists of a distributor plate with horizontal slots and an impingement plate mounted in front of the distributor plate, see Fig.\ref{fig:Top_Section}. 
The height and width of the slots were determined through iterative preliminary testing, aimed at optimizing the homogeneity of the film flow. The optimal configuration found is composed of four slots spanning the entire channel width, with a progressively increasing height from the bottom to the top. The use of the impingement plate ensures that the liquid is directed in the main flow direction. This results in a further improvement in homogeneity. 
The generated film, then flows downwards into the upper channel section, illustrated in Fig \ref{fig:Channel}. Within the upper channel section qualitative and quantitative investigations of the falling film flow takes place under controlled conditions. It is composed of perspex forming the lateral and frontal walls, while a copper plate constitutes the rear surface over which the film flows. Due to the use of transparent perspex, accessibility is ensured for lateral and frontal optical investigations. Copper is used as rear wall material due to its high thermal conductivity.
Therefore, promoting a uniform temperature distribution across the plate, which is necessary for future investigations into falling film evaporation. The copper plate is connected to an aluminium plate. Thermal conductive paste was applied in between the plates to ensure an optimal heat transfer. The aluminium plate is equipped with three individually controllable heating pads. The heating pads can be used to adjust the surface temperature of the copper in case of evaporation experiments. The wall temperature regulation will be realized by individually controlling the heat flux of each heating pad via a PID-controlled power supply. Thus, the wall temperature of the copper must be measured. For this purpose type K thermocouples are placed within blind holes reaching up to 1mm below the copper surface exposed to the falling film.

After leaving the upper channel section the film flows through the lower channel section. This section ensures, that the uprising gas flow is hydrodynamically fully developed before entering the area where investigations of the film flow under controlled conditions are desired. At the bottom of the lower channel section, the liquid enters the plenum. Within the plenum the liquid is stored during the experiments. Additionally, the gas gets introduced into the plenum, thus gas pressure fluctuations are equalized allowing a more even distribution during experiments. To further ensure an uniform gas inflow from the plenum into the lower channel section, both parts are interconnected with a nozzle. The nozzle also  contains a socket for the honeycomb and the turbulence grids at its inlet. Thus allowing to control the inlet turbulence intensity in the channel.

\subsection{\label{sec:level2}Measurement devices and methods}

\begin{figure}[h!]
  \includegraphics[width=1\columnwidth]{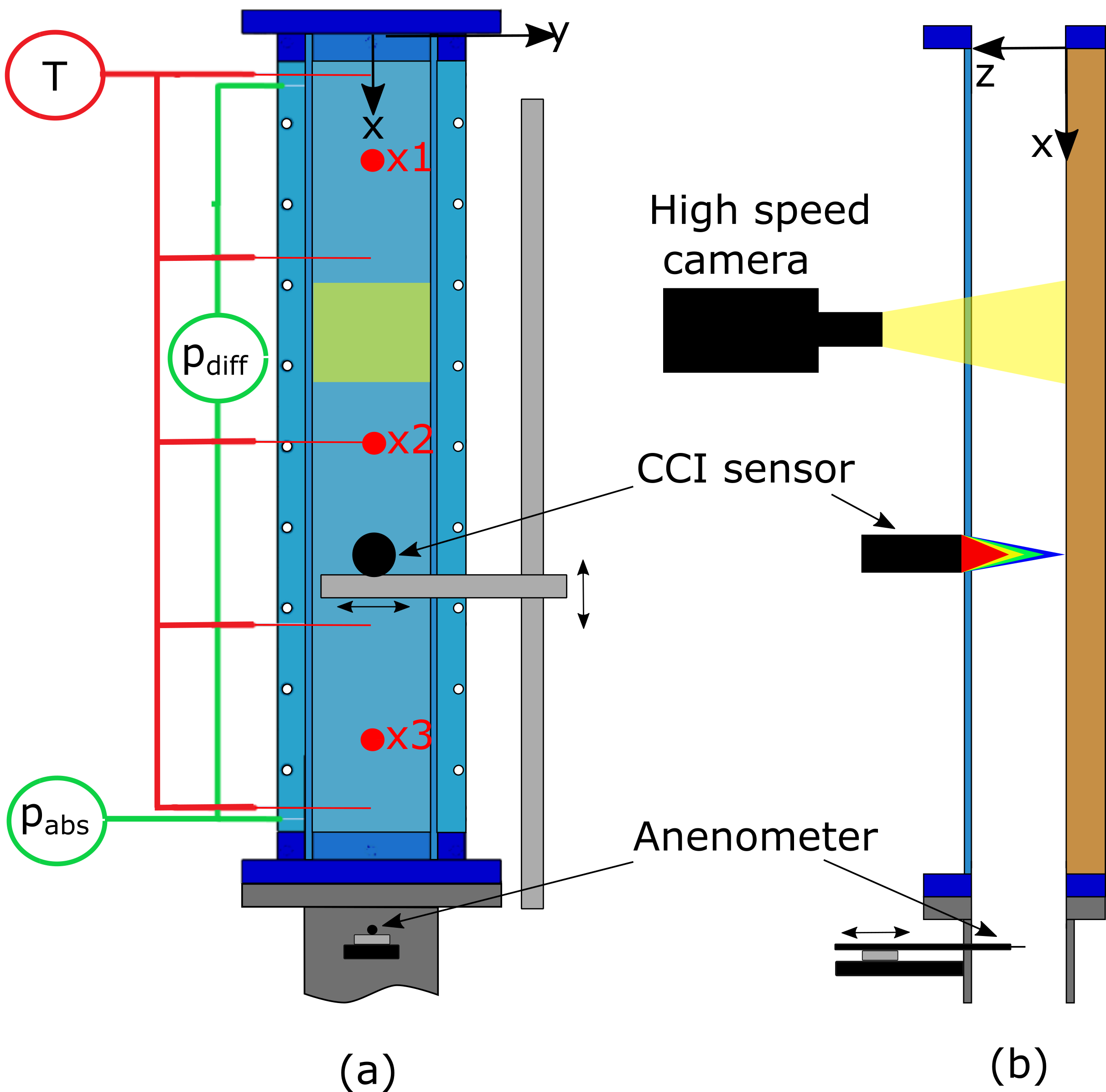}
  \caption{Schematic representation of measurement devices and measurement positions}
  \label{fig:Measurement_devices}
\end{figure}

To investigate the occurring flow phenomena within the measurement section qualitative and quantitative measurement methods are applied. An overview of the utilized measurement devices is illustrated in Fig. \ref{fig:Measurement_devices}.
For the qualitative studies of the film flow, high speed camera imaging was utilized. For this purpose, a Kron Technologies Chronos 1.4 RGB high speed camera with a maximum framerate of 1069 frames per second, was mounted in front of the upper channel section. The recording area covers the entire channel width of 100mm and enables the visualization of film flow patterns. Thus,  qualitative macroscopic flow phenomena can be linked with quantitative microscopic examinations of the film made with a Confocal Chromatic Imageing sensor (CCI). The CCI sensor operates on the principle of chromatic abbreviation of a polychromatic light and allows temporal resolved point wise measurements of the film thickness. The advantages of CCI sensors include their non-invasive measurement principle, high accuracy and high temporal resolution as well as insensitivity to environmental influences. 
The working principle is explained in detail e.g. by Schr\"oder \cite{tuprints8742} and was used in several investigations of falling film flows \cite{KOFMAN201722,HUANG2014125,ZHOU2009273}. For film thickness measurement on FETRIG, a KEYENNCE CL-L030 measuring head is utilized, allowing film thickness measurements with a resolution of  0.25 \( \mu m\). The CCI-Sensor is integrated into a 2-axis CNC system, allowing fast, remote controlled positioning in front of the upper channel section. The CNC system further prevents temperature and humidity fluctuations during measurements, as it allows for the positioning of the sensor without the necessity of entering the climatic chamber. In the initial measurement campaign, the film thickness measurements with the CCI sensor were conducted at the positions x1 = 100 mm, x2 = 350 mm, and x3 = 600 mm, as illustrated in Fig.\ref{fig:Measurement_devices}.
For the gas phase quantitatve measurements of differential and absolute pressure, temperature as well as the velocity distribtution within the gas phase are conducted. To determine the differential and absolute pressure the corresponding sensors are connected via hoses to two pressure measuring orifices. The orifices are located on the top and bottom of the left side wall of the upper channel section. Thus, the pressure loss across the upper channel section is measured via a differential pressure sensor.  For this purpose an OMEGA PXM409-025HDWU10V with a measurement range of 0 to 25 mbar is utilized. The absolute pressure measurements are conducted via an OMEGA PXM409-002BA10V with a measurement range of 0 to 2 bar at the bottom orifice. For temperature measurements five type K thermocouples are inserted through orifices at the same side wall with an equidistant distance. All thermocouples are placed in such a way that their measuring tips are located in the center of the corresponding duct cross-section. 
To investigate the gas velocity close to the gas inlet of the upper channel section a 10 \(\mu m\) hot wire anemometer, provided by SVMtec has been incorporated. The anemometer is capable of gas velocity measurements up to 45 m/s in upstream direction. The anemometer and its attachments are integrated at the top of the lower channel section to prevent any obstruction in the visible area in the perspex front wall. To enable the measurement of velocity distributions, the anemometer is mounted on a traversing unit centered on the front of the channel. This traversing unit allows the remote-controlled positioning of the anemometer within the channel in horizontal direction. As a result, velocity profiles can be determined along the centerline of the channel's cross-section, perpendicular to the wetted surface.

\section{\label{sec:level1}Results}
The purpose of constructing FETRIG was to establish a facility to investigate the flow  phenomena occurring in TCPTs under controlled and well-defined conditions. To ensure that the system fulfills the intended requirements, the functionality and limitations of FETRIG must be investigated.  The performance of FETRIG and the suitability of the choosen measurement systems for hydrodynamic investigations were evaluated in an preliminary measurement campaign. The selected fluids and operational parameters are listen in Tab. \ref{tab:table1}. Water was used due to its similar Kapitza number compared to  $CO_{2}$ used in the corresponding TCPT, see Hagedorn et al.\cite{Hagedorn_2024}. As there is no safety risk due to the use of water, the gas circuit was operated in the open configuration mode. Thus, conditioned humid air within the climate chamber was used as gas.

\begin{table}[h!]
\caption{\label{tab:table1}Used fluids and operating parameters}
\begin{ruledtabular}
\begin{tabular}{ll}
Parameters& Settings \\
\hline
Liquid & Distilled Water \\
Gas & Air\\
Temperature  & 298.15 K\\
Relative Humidity & 80 \%\\
Gas Cycle mode & Open Configuration\\
$Re_{liq}$& 0;500;620;740;860;980;1,100\\
$Re_{gas}$& 0;4,900;8,900;12,900;16,900;20,900\\
Inclination & 90° \\
Ka & $5.4 \times 10^{10}$

\end{tabular}
\end{ruledtabular}
\end{table}
A minimum stable adjustable gas flow is established for  \(Re_{gas}\) > 4,900, limited by the measurement accuracy of the LFE. The maximum gas flow was restricted by the blower to  \(Re_{gas}\) < 22,900. The liquid flow was limited to  \(Re_{liq}\) > 500 and \(Re_{liq}\) < 1,100.  Preliminary tests indicated a minimum detectable film thickness of approximately 200 \(\mu m\)  using the CCI-sensor in case of water. At \(Re_{liq}\) < 500,  the film thickness experienced progressive undercutting of this limit, resulting in significant challenges for accurate film thickness measurements. 
The upper limit was determined by performance of the liquid pump as well as the calibrated measurement range of MFM2. 
\subsection{\label{sec:level2}Characterisation of single gas and liquid flow}
\begin{figure}[!ht]             % Einbetten in figure wie gehabt
\centering                  % zentrierte Ausrichtung, optional
\def\svgwidth{270pt}    
% die Bildbreite muss auf diese Weise festgelegt werden!
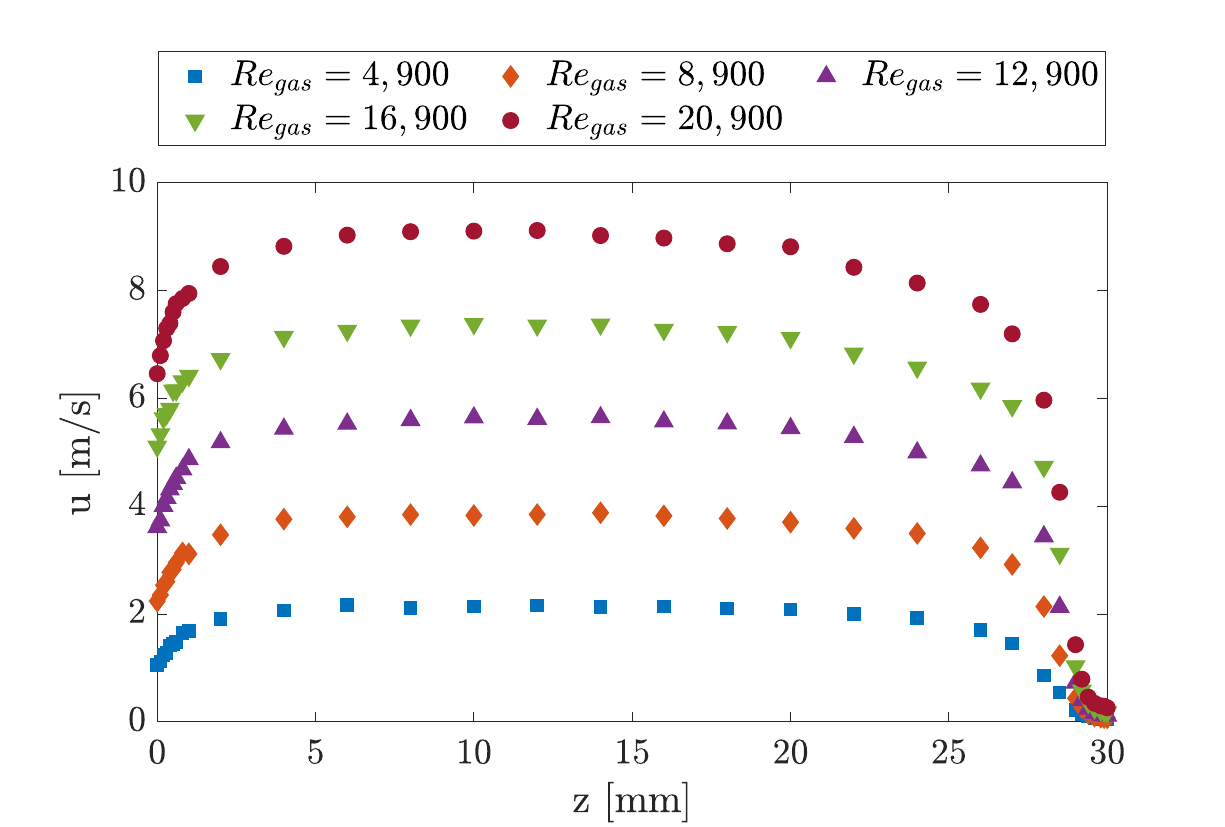  
% 'versuchsaufbau' durch Dateinamen ersetzen.
\caption{Air velocity profiles measured without liquid film at the top of the lower channel section.}    % Bildunterschrift, optional
\label{fig:Velocity_Profile_Gasphase}          % Label f�r Verweise, optional
\end{figure}

\begin{figure}[h!]
  \includegraphics[width=1\columnwidth]{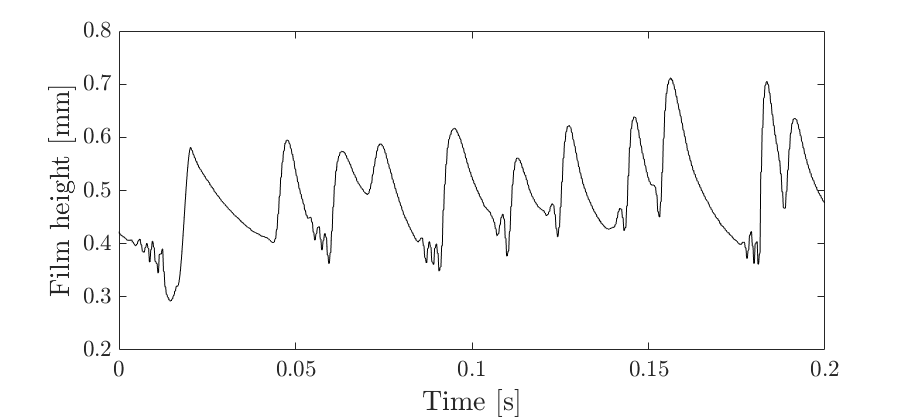}
  \caption{Time depended film thickness measurement data at x = 350 mm with \(Re_{liq}=500\), \(Re_{gas}= 0\) with a sampling rate of 5,000 Hz. }
  \label{fig:thickness_profile}
\end{figure}

\begin{figure}[!ht]             % Einbetten in figure wie gehabt
\centering                  % zentrierte Ausrichtung, optional
\def\svgwidth{270pt}    
% die Bildbreite muss auf diese Weise festgelegt werden!
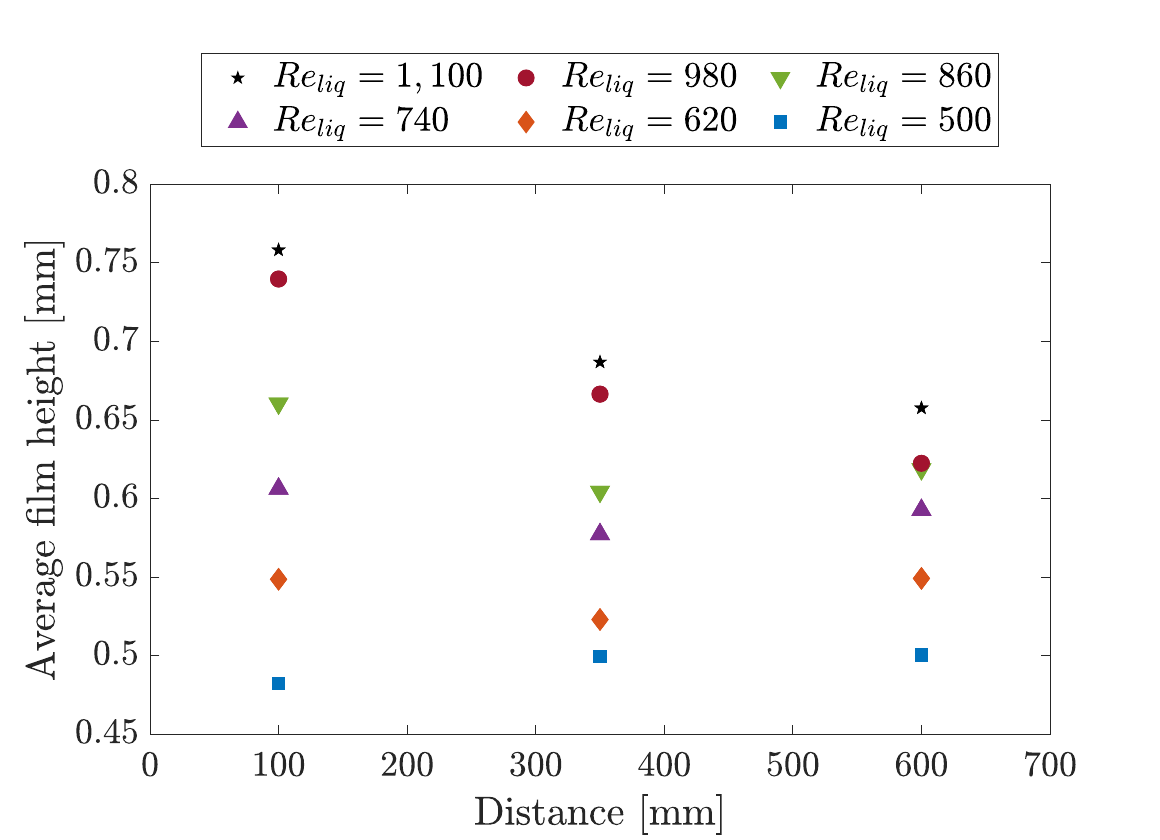  
% 'versuchsaufbau' durch Dateinamen ersetzen.
\caption{Time averaged film thickness development in stream wise direction, measured along the plate at x = 100 mm, 350 mm, 600 mm within the upper channel section for different \(Re_{liq}\) and \(Re_{gas}=0\).}    % Bildunterschrift, optional
\label{fig:ReG0_ReLVar_XVar}          % Label f�r Verweise, optional
\end{figure}

The velocity profiles measured with the hot wire anemometer prior entering the upper channel section are shown Fig. \ref{fig:Velocity_Profile_Gasphase}.  Due to the fragility of the 10 \(\mu m\) hot wire, no film was employed to prevent destruction.  The measurements were conducted for all in  Tab. \ref{tab:table1} listed \(Re_{gas}\). Each dot corresponds to a time-averaged measurement over \(40\) s to reduce temporal fluctuations.
It is important to note that position z = 0 does not correspond to the opposite channel wall but but rather to the closest measurement position, see Fig. \ref{fig:Measurement_devices}. This limitation arised from the fragility of the hot wire tip, which necessitates a minimum distance to avoid damaging.
As a result, the distance to the wall could not be determined directly, resulting in a measurement of velocities of u(z=0) > 0.
CCI sensors were already successfully utilized in investigations of falling film flows with counter-current gas, as demonstrated by Kofman and Mergui\cite{KOFMAN201722} and Huang et al. \cite{HUANG2014125}. However, it is necessary to validate the applicability of this measurement system for FETRIG. It must be ensured that the film thickness measurement range and resolution, as well as the temporal resolution are capturing all appearing flow phenomena. A sequence of a measured temporal film thickness signal is shown in Fig. \ref{fig:thickness_profile}. 
The measurements were obtained at  x = 350 for \(Re_{liq}\) = 500 without gas flow and a sampling rate of 5,000 Hz. The sampled data show two clearly distinguishable wave structures. Low-amplitude, high-frequency capillary waves followed by high-amplitude low-frequency solitary waves. This wave pattern is typical for falling film flows, and the qualitative results are in agreement with experiments conducted by Kofman and Mergui \cite{KOFMAN201722}. A sampling rate of 5,000 Hz was determined as optimal across all operating points. This condition is attributed to the inability of lower sampling rates to effectively capture high-frequency capillary waves. In contrast, increasing the sampling rates beyond this point does not yield any significant additional information but instead amplifies noise levels, resulting in an increased proportion of invalid data. The sampled film thickness data were used to determine the average film height $\bar{h}_{film}$, enabling  a time-independent interpretation. Figure \ref{fig:ReG0_ReLVar_XVar} displays the average film height evolution in main flow direction captured for all investigated \(Re_{liq}\) and measurement position. 
With increasing \(Re_{liq}\) an increase in $\bar{h}_{film}$ is visible, thus showing the expected trend. In flow direction however, the film thickness increases slightly for \(Re_{liq}=500\) and decreases for \(Re_{liq}\) = 980 and \(Re_{liq}\) = 1,100, while no clear trend can be distinguished for \(Re_{liq}\)= 620, \(Re_{liq}\) = 740 and \(Re_{liq}\) = 860. The increase in film thickness for \(Re_{liq}\) = 500 indicates a downstream deceleration of the film towards the equilibrium state, as described by St\"ucheli and \"Ozisik \cite{STUCHELI1976369}. In contrast to this, for high \(Re_{liq}\) further acceleration downstream might cause the decrease in film thickness. Accordingly, this suggests that a hydrodynamically developed film can be assumed for the film flows without a distinct trend. To provide a more accurate assessment of the film thickness variation in the flow direction, further investigations with enhanced spatial resolution are planned.
\subsection{\label{sec:level2}Characterisation of counter current liquid film gas flow}
\begin{figure}[!ht]             % Einbetten in figure wie gehabt
\centering                  % zentrierte Ausrichtung, optional
\def\svgwidth{270pt}    
% die Bildbreite muss auf diese Weise festgelegt werden!
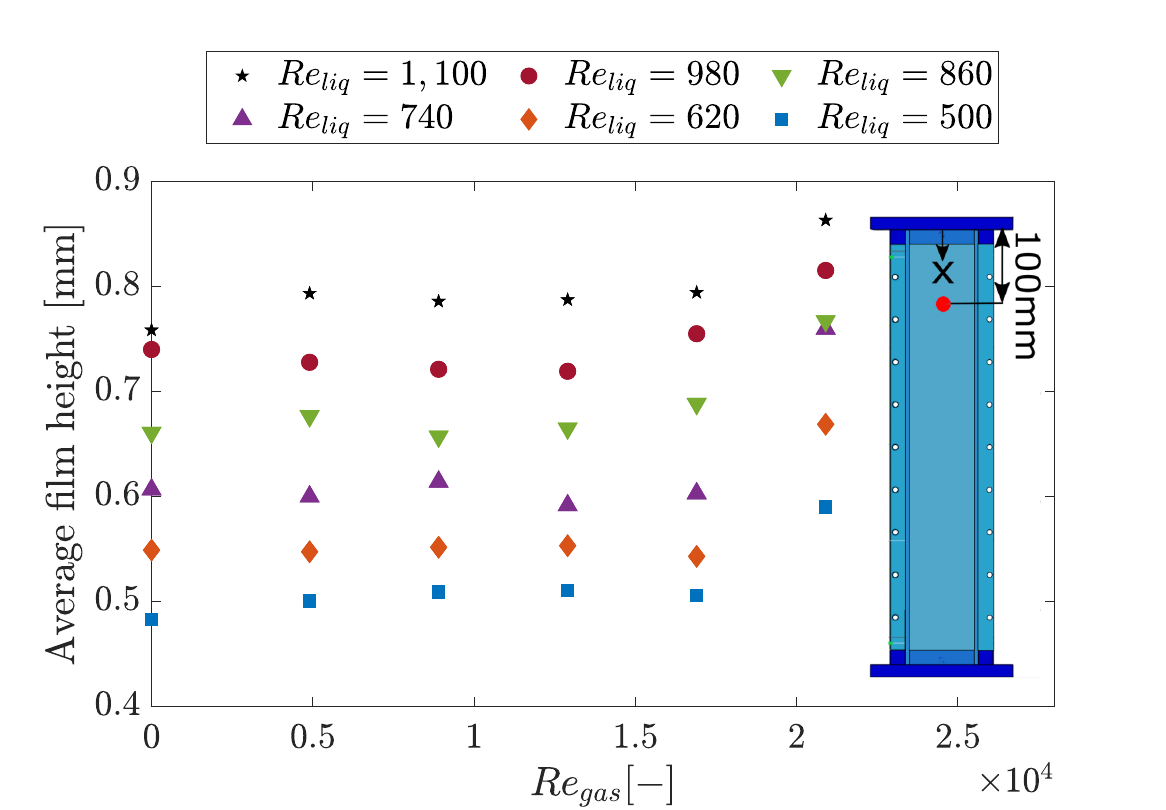  
% 'versuchsaufbau' durch Dateinamen ersetzen.
\caption{Influence of the gas flow on the time averaged film thickness at x = 100 mm for different \(Re_{liq}\)  and \(Re_{gas}\).}    % Bildunterschrift, optional
\label{fig:ReGVar_ReLVar_X100}          % Label f�r Verweise, optional
\end{figure}

\begin{figure}[!ht]             % Einbetten in figure wie gehabt
\centering                  % zentrierte Ausrichtung, optional
\def\svgwidth{270pt}    
% die Bildbreite muss auf diese Weise festgelegt werden!
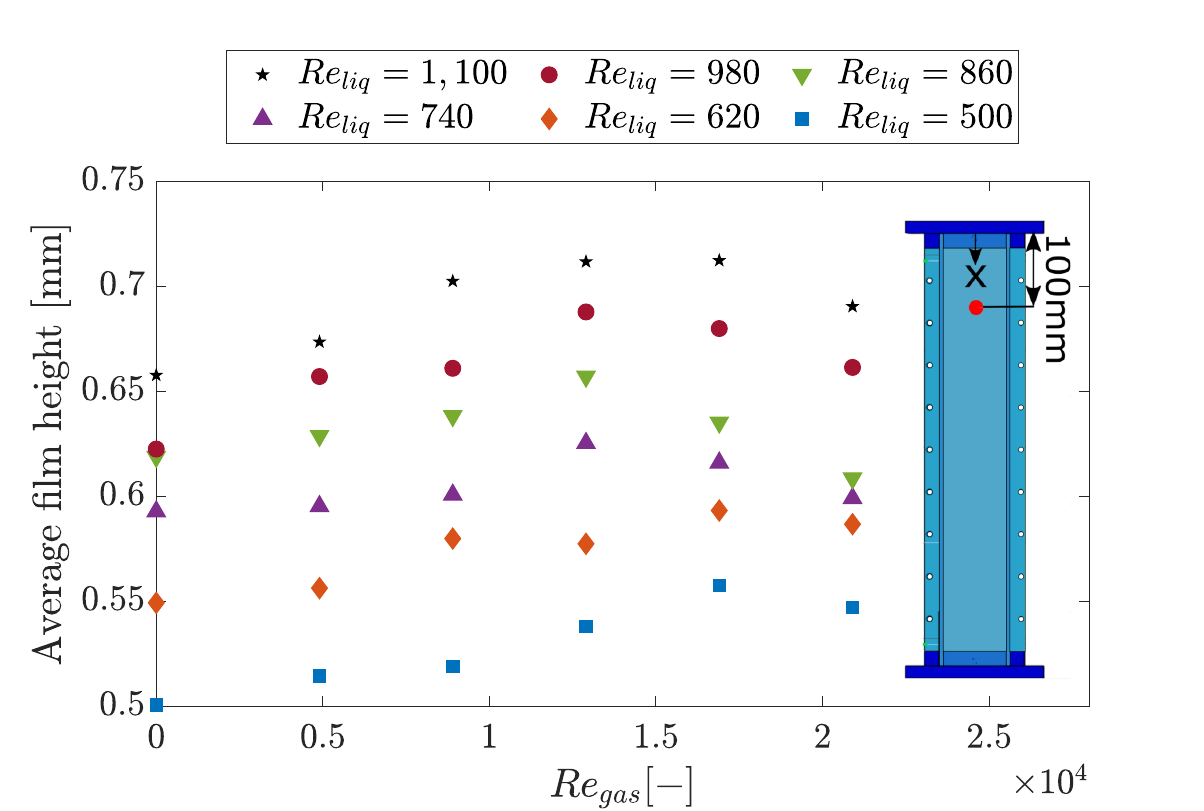  
% 'versuchsaufbau' durch Dateinamen ersetzen.
\caption{Influence of the gas flow on the time averaged film thickness at x = 600 mm for different \(Re_{liq}\)  and \(Re_{gas}\).}    % Bildunterschrift, optional
\label{fig:ReGVar_ReLVar_X600}          % Label f�r Verweise, optional
\end{figure}

The study of the influence of \(Re_{liq}\) on  $\bar{h}_{film}$ was done for all  \(Re_{liq}\) and  \(Re_{gas}\) listed in Tab. \ref{tab:table1} as well as for all measurement positions of the CCI sensor shown in Fig. \ref{fig:Measurement_devices}. Due to the similiarity inbetween the measured values at position  x = 350 mm and x = 600 mm only the values of the latter one will be presented. In Fig. \ref{fig:ReGVar_ReLVar_X100} the variation of film thickness with superficial gas velocity is presented at x = 100.
It is evident that an increase in \(Re_{gas}\) does not significantly affect $\bar{h}_{film}$ below a threshold \(Re_{gas}\), beyond which an increase in $\bar{h}_{film}$ is visible for all  values of \(Re_{gas}\). The phenomenon of film thickening at higher \(Re_{gas}\), and consequently higher gas velocities, has also been described by Al Sayegh et al. \cite{SAYEGH2022104170}. Moreover, the threshold value tends to decrease with increasing \(Re_{liq}\). For  \(Re_{liq}\) < 740, an increase in $\bar{h}_{film}$ is not visible for  \(Re_{gas}\) < 16,900, whereas  for \(Re_{liq} \geq \) 740, an increase of  $\bar{h}_{film}$ becomes apparent at \(Re_{gas}\) = 12,900.
The influence of the gas flow on $\bar{h}_{film}$ at x = 600 mm  is shown in Fig.\ref{fig:ReGVar_ReLVar_X600}. In contrast to x = 100 mm, an increase in $\bar{h}_{film}$ is already distinguishable at the lowest \(Re_{gas}\). Furthermore, an opposite trend is observed onwards the threshold values, as $\bar{h}_{film}$ decreases beyond this point. 
Comparable tendencies were also reported by Huang et al. \cite{HUANG2014125} and Drossos et al. \cite{DROSOS200651}, who attributed it to the onset of flooding. According to Huang et al. \cite{HUANG2014125}, the decrease in film thickness can be explained by a decrease of the liquid mass flow within the film, caused by entrainment into the air flow.  In addition to film thickness measurements,  visual observations of the film flow can also be employed as a method for detecting flooding. According to Drossos et al. \cite{DROSOS200651}, in case of flooding a disintegration of the coherent wave pattern can be observed. The influence of the gas flow on the liquid film structure is illustrated in Fig.\ref{fig:Morphology_film_ReG0_ReG20,900}. 
\begin{figure}[h!]
  \includegraphics[width=1\columnwidth]{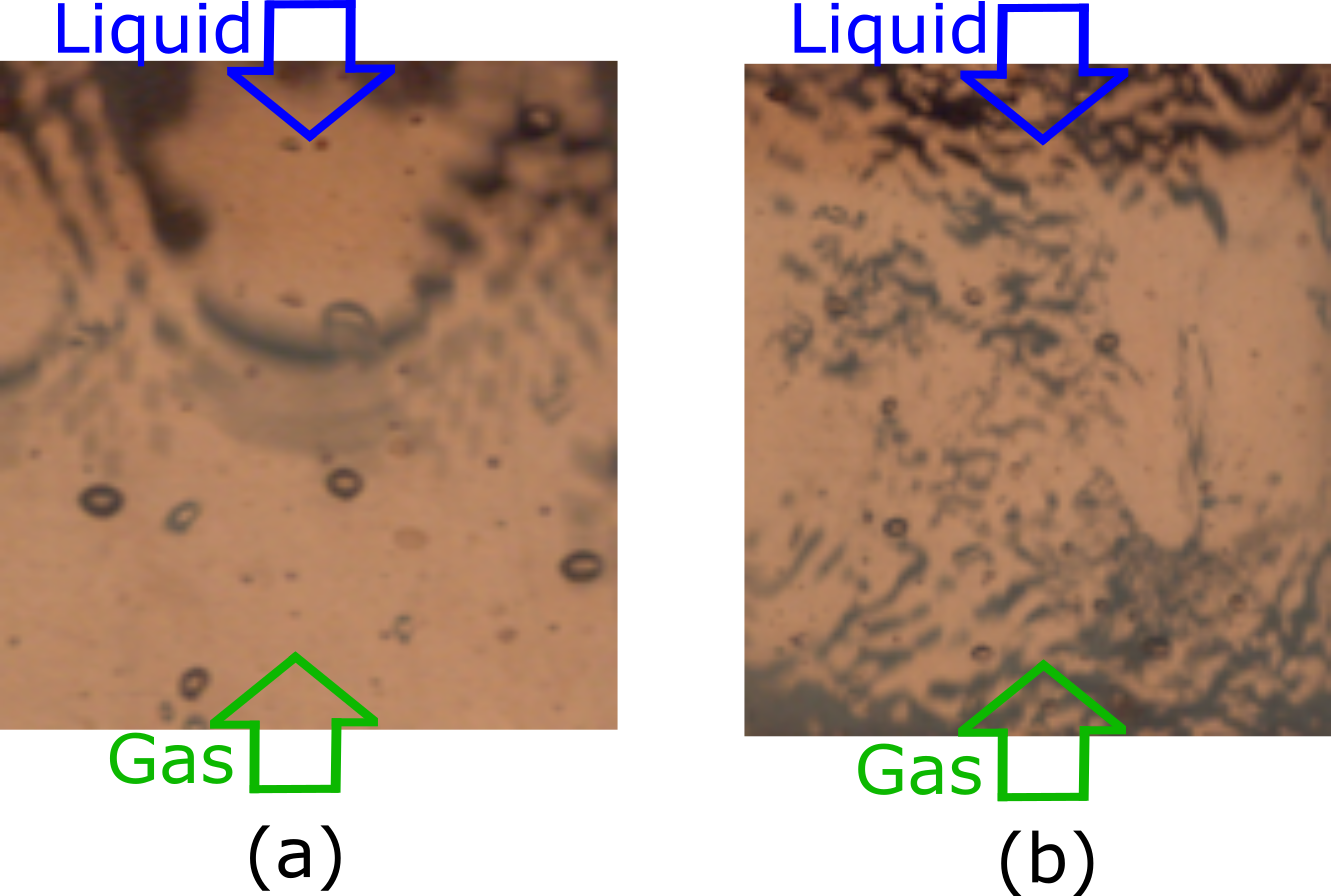}
  \caption{Effect of the gas velocity on liquid film flow for \(Re_{liq}\) = 500 at x = 250 with (a) \(Re_{gas}\) = 0 and (b) \(Re_{gas}\) = 20,900.}
  \label{fig:Morphology_film_ReG0_ReG20,900}
\end{figure}
Figure\ref{fig:Morphology_film_ReG0_ReG20,900}(a) presents a photograph of a horse-shoe-shaped wave, preceded by capillary waves, at \(Re_{liq}\) = 500 without gas flow. In contrast, Fig.\ref{fig:Morphology_film_ReG0_ReG20,900}(b) shows a highly chaotic pattern at the same \(Re_{liq}\) and position with \(Re_{gas}\) = 20,900. This reinforces the assumption that flooding occurred. Nevertheless, during the measurements, flooding was not observed within the upper channel section, which  indicates that flooding occurred within the top section. 
\section{\label{sec:level1}Conclusion}
In this paper, the design of a new test rig for falling film investigations is presented, and the measurement methods employed are explained. Due to the lack of models for describing the flow phenomena occurring TCPTs, their design and optimization are limited. To validate future models, the test rig, called FETRIG, has been developed with high requirements for precision and reproducibility.

The design of FETRIG enables the investigation of falling film flows with counter-current gas flow under highly controlled conditions, with independent control of \(Re_{liq}\) and \(Re_{gas}\). The inclinable measurement section is located within a climate control chamber, allowing variations of the fluids material properties and ensuring constant conditions during measurements, as well as repeatability between experiments. Characterized by its modular design and optical accessibility, the measurement section facilitates the straightforward application of various measuring devices. Both qualitative and quantitative measurement methods have been employed. For qualitative investigation   of the flow patterns of the falling film, a high-speed RGB camera was utilized. A CCI sensor was used for non-invasive, temporally resolved, pointwise measurements of the film thickness. The CCI sensor was selected for its high temporal resolution and accuracy, as well as its insensitivity to environmental influences. A hot-wire anemometer has been implemented to measure the velocity profiles within the gas phase.
The performance of the test rig and the applicability of the measurement systems have been evaluated during a measurement campaign. For this investigation, water was employed as the liquid, while air with a relative humidity of 80\% was used as the gas, both maintained at a temperature of 298.15K and an inclination angle of 90° of the measurement section. The velocity profiles of the gas phase were measured within the range of 4,900 < \(Re_{gas}\) < 20,900, without the presence of liquid. Film thickness measurements were performed within the range of 500 < \(Re_{liq}\) < 1100. The observation of typical wave patterns occurring in falling film flows confirmed that the measurement range, resolution, and temporal resolution of the CCI sensor are suitable for investigating falling film flows. Furthermore, it was demonstrated that measurements of film thickness can be obtained in the presence of counter-current gas flow.

The developed facility provides an accurate tool for investigating falling films. Initial experiments examining both gas flow and film flow have been successfully conducted. 

\begin{acknowledgments}
The authors acknowledge the financial support of this work from the Deutsche Forschungsgemeinschaft (DFG) under project no. 455199958. 

The following article has been submitted to Review of Scientific Instruments (RSI). After it is published, it will be found at \href{https://pubs.aip.org/aip/rsi}{Link}.
Copyright (2025) M. Wirth, J. Hagedorn, B. Weigand, S. Kabelac. This article is distributed under a Creative Commons Attribution-NonCommercial 4.0 International (CC BY-NC) License. \href{https://creativecommons.org/licenses/by-nc/4.0/}{Link}

\end{acknowledgments}

\section*{Author Declarations}
\subsection*{Conflict of Interest}
The authors have no conflicts to disclosure.

\section*{Data Availability Statement}

The data that support the findings of this study are available from the corresponding author upon reasonable request.

%\printbibliography %Prints bibliography

\section*{References}
\nocite{*}
\bibliography{mwi_bibliography}% Produces the bibliography via BibTeX.

\end{document}